# Efficient single-emitter plasmonic patch antenna fabrication by novel deterministic in situ optical lithography using spatially modulated light

Amit Raj Dhawan[1,2,4], Michel Nasilowski[3], Zhiming Wang[1], Benoît Dubertret[3], Agnès Maître[2*]

[1]Institute of Fundamental and Frontier Sciences, University of Electronic Science and Technology of China, Chengdu 610054, People's Republic of China

[2]Sorbonne Universités, UPMC Univ Paris 06, UMR 7588, Institut de NanoSciences de Paris (INSP), Paris F-75005, France

[3]Laboratoire de Physique et d'Etude des Matériaux, ESPCI-ParisTech, PSL Research University, Sorbonne Université UPMC Univ Paris 06, CNRS, 10 rue Vauquelin 75005 Paris, France

[4] Department of Materials, University of Oxford, Parks Road, Oxford, OX1 3PH, UK

***Corresponding author**. Email: agnes.maitre@insp.upmc.fr, Telephone: +33 1 44 27 42 172

**Running title**: Dhawan *et al*, Deterministic lithography with donut laser modes

*Single-emitter plasmonic patch antennas are room-temperature deterministic single photon sources, which exhibit highly accelerated and directed single photon emission. However, for efficient operation these structures require three-dimensional nanoscale deterministic control of emitter positioning within the device, which is a demanding task, esp. when emitter damage during fabrication is a major concern. To overcome this limitation, our deterministic room-temperature in situ optical lithography protocol uses spatially modulated light to position a plasmonic structure non-destructively on any selected single-emitter with three-dimensional nanoscale control. In this paper we analyze the emission statistics of such plasmonic antennas that embed a deterministically positioned single colloidal CdSe/CdS quantum dot that highlight acceleration and brightness of emission. We demonstrate that the antenna induces a 1000-fold increase in the emitter absorption cross-section, and under high pumping, these antennas show nonlinearly enhanced emission.*

# Introduction

The possibility of deterministically controlling the local environment of an emitter to make its emission faster, the capability of collecting more emitted photons, and the effectiveness of exciting a system with only one photon will benefit several optical quantum technologies [1]. The highly optimized light–matter interactions in devices with these features will highlight the involved physics [2] and find application in fields like single photon generation [3,4,5], quantum plasmonics [6, 7], and optical antennas [8, 9, 10]. Advances in microlithography [11] produced superior electronics and helped attain the first quantum revolution, and the developments in nanofabrication [12] are paving the way for the second quantum revolution [13].

By means of Purcell effect [14], structures such as optical cavities [15] and plasmonic antennas [16, 17], modify light emission. Their optimal operation necessitates precise spatial positioning of the emitter with respect to its environment, which is a challenging task. Although electron-beam lithography obtains higher resolution than optical lithography, it requires more expensive and specialized equipment and working conditions. Moreover, the beam electrons can damage the emitter. Optical lithography is a cheaper and convenient solution that achieves good writing resolution without electron exposure. However, the intense light required to perform lithography locally above the emitter can photodegrade it. Our fabrication protocol [18] resolves this problem by selectively not exposing the emitter to intense light during the lithography as it uses a custom-designed donut laser beam. At room-temperature and under ambient working conditions, we have employed this protocol to position a plasmonic patch on any selected CdSe/CdS QD with nanometric vertical and horizontal precision, and fabricate plasmonic patch antennas with remarkable emission properties.

Core/shell QDs [19], vacancy centers in nanodiamonds [20], and defects in two-dimensional materials [21, 22] are deterministic room-temperature single photon sources. Coupling them to optical or plasmonic cavities can make their emission brighter and faster. Optical cavities require very precise spectral matching, which makes them unsuitable for broadband emitters like colloidal QDs at room-temperature. However, such broadband emitters can couple very efficiently to plasmonic antennas due to their wide spectral resonance and low volume, thus making them promising room-temperature single photon sources. This has been discussed [17,23,24] and demonstrated [25,26] recently. Embedding the QD between the nano-spaced metal-dielectric interfaces couples its radiation to generate surface plasmon polaritons (SPPs) at the interfaces [17], which creates very high electromagnetic field around the QD and accelerates its emission. As the patch thickness is less than the skin depth of the field, the SPP-coupled interaction in the antenna emits photons from the patch as shown in Figure 1(a).

Here we describe the deterministic fabrication of single-emitter patch nanoantennas and demonstrate that the antenna can dramatically modify QD emission, which is illustrated by high recombination rate and fluorescence enhancement, increased absorption cross-section, and the nonlinear emission of the antenna. The metallic losses in plasmonic antennas that reduces their radiative efficiency [27] can be mitigated by their acceleration and directionality of emission, thus achieving faster and brighter emission. We show that a plasmonic antenna can increase the absorption cross-section of a QD by more than three orders, and lead to highly nonlinear emission under low power excitation.

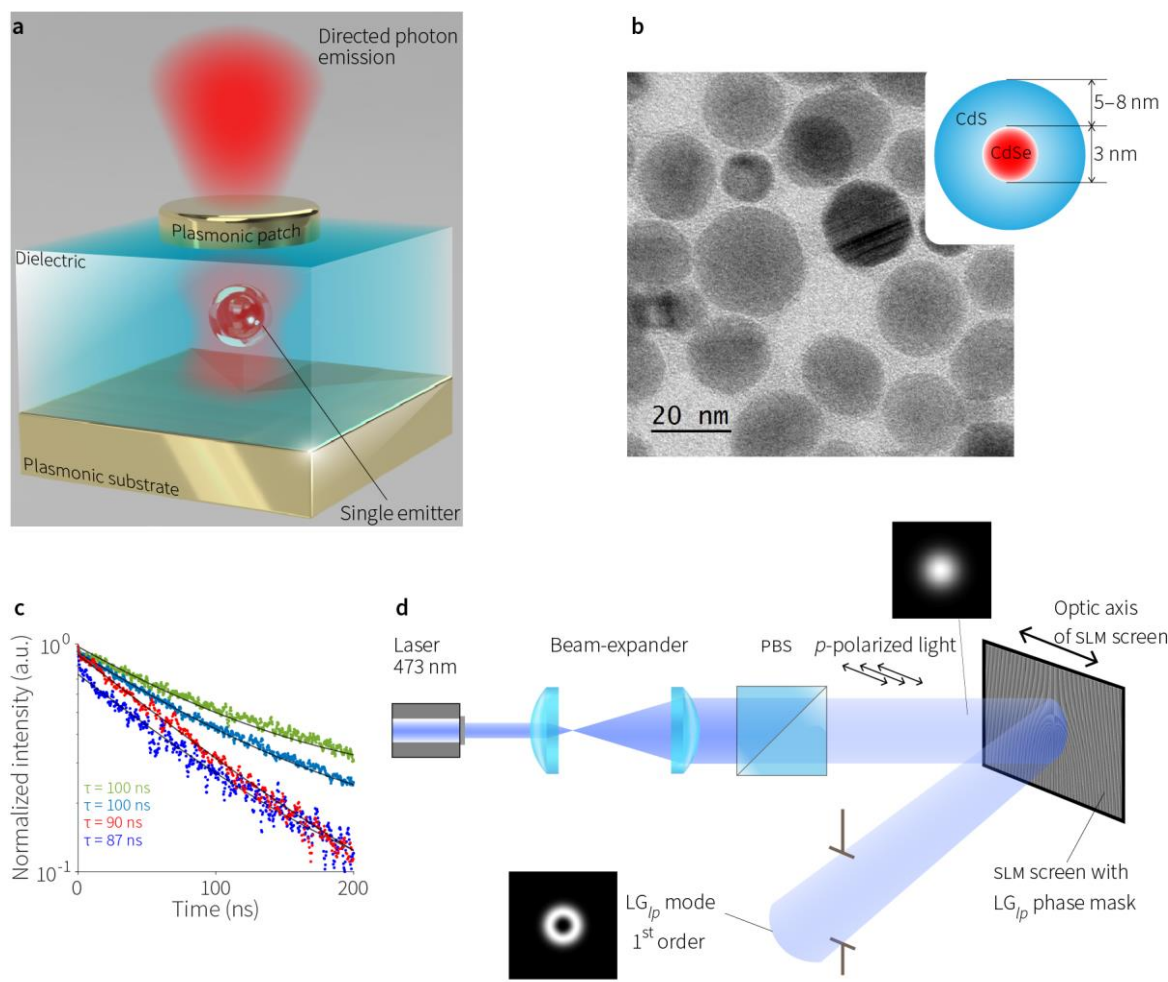

**Figure 1** *Single-emitter plasmonic patch antenna with a QD*

A single QD is selected and embedded in a dielectric layer sandwiched between a plasmonic metal film and a patch to obtain a single-emitter plasmonic patch antenna, as illustrated in **a**. A transmission electron microscopy image (**b**) of the colloidal CdSe/CdS QDs used in this work, and **c**, monoexponentially fitted emission decays of four QDs in a homogeneous medium. **d** Optical setup used for the generation of spatially modulated light using a reflective phase-only spatial light modulator. PBS is a polarizing beam splitter.

## Results

**Deterministic in situ spatially modulated laser lithography**

Conventional photolithography [28] utilizes light of appropriate wavelength and intensity to locally expose the photoresist, which is then chemically etched. Using in situ photolithography, Belacel *et al.* [10] demonstrated deterministic fabrication of plasmonic patch antennas that contain an aggregate of colloidal QDs. This method cannot be extended straightforwardly to room-temperature deterministic fabrication of single-emitter antennas because the fluorescence of a single QD is considerably lower than that of an aggregate, and therefore it is masked by the luminescence of the photoresist. The high light intensity required to observe a QD leads to unintended exposure (chemical modification due to light) of the photoresist above it and makes localized lithography impossible. Our protocol avoids this problem by using a resist bi-layer which has low luminescence and is not exposed like conventional photoresists during the process. The bi-layer resist stack consisting of a lift-off resist (LOR® [29]) and polymethyl methacrylate (PMMA) does not rely upon conventional exposure because it is evaporated by intense laser light during the laser etching process [30]. However, the intense light of the fundamental laser mode required to remove the resist bi-layer above a single QD usually results in photobleaching it. We resolve this problem by using spatially modulated laser light with a donut profile (Figure 1d), which does not expose the QD to light during the lithography, and the carefully designed intensity profile of the laser mode leads to a successful positioning a plasmonic patch centered/off-centered above the QD with a lateral precision of ±50 nm. The QD can be positioned vertically in the antenna with a precision of ±3 nm using spin-coated thin films. This lithography protocol can work with a variety of emitters—single or aggregates—such as QDs, vacancy centers in nanodiamond, molecules, and defects in two-dimensional materials. This article describes the use of this protocol to deterministically select and position a single colloidal CdSe/CdS QD (Figure 1b) inside a plasmonic patch antenna as illustrated in Figure 1a. Light–matter interactions in nanophotonic devices can be optimized by deterministic control over emitter selection and device fabrication, and the described lithography protocol is capable of achieving this. The antennas discussed in this paper were fabricated at room-temperature and under atmospheric pressure, but the fabrication protocol can be used at low temperature and in vacuum conditions.

Compared to our other protocol [25] that utilizes a laser with wavelength tunability to maximize absorption of light by the resist and minimize QD photobleaching, this method excels at preventing emitter damage as no light is seen by the QD during the laser etching step. Moreover, by modifying the phase map on the spatial light modulator (SLM) screen, a variety of high-resolution shapes and patterns can be created. The confocal scan and the laser etching pattern of Figures 2a, b were produced by a Laguerre-Gaussian (LG) donut mode (here $LG_{40}$) that was focused through a microscope objective. The mode was created by reflecting a fundamental mode laser beam using an optimally tuned liquid crystal SLM displaying a phase mask, and then filtering out the ±1 diffraction order as shown in Figure 1d.

On a 280 µm thick Si wafer, a 10 nm thick adhesive layer of Cr is deposited, above which a 200 nm thick optically opaque layer of Au is evaporated. Then a 10 nm thin layer of PMMA is spin-coated and baked, on which a QD dispersion of adequate concentration is spin-coated to obtain well-separated single QDs. Then another layer of PMMA (35 nm) is spin-coated and baked, thus embedding individual QDs in a dielectric layer. Then a lithography resist bi-layer is created by firstly spin-coating and baking a 450 nm thick layer of LOR®5A, and then a 50 nm thick layer of PMMA.

The sample (Figure 2a) is confocally scanned using low intensity laser light (473 nm continuous wave laser). After positioning the emitter, high intensity laser light is used to remove the resist bi-layer locally above the emitter as shown in the atomic force image (AFM) above Figure 2b. As the light intensity of the laser mode is almost zero at the center as seen in Figure 2f, the resist at the center is not removed and the QD below it does not photodegrade during the lithography. Then by selective chemical etching of LOR®5A, the remaining central cylinder of the resist bi-layer is removed, and an undercut, which facilitates the final lift-off, is created in the upper PMMA (Figure 2c). The size and shape of the etched feature can be controlled precisely by modifying the laser beam profile and the etching time or chemicals. After evaporating 20 nm of gold (Figure 2d) on the sample, a chemical lift-off is performed to obtain single-emitter patch antennas (Figure 2e).

The performance of these antennas depends on the dipolar orientation of the QD, the dielectric spacer thickness around the QD, the position of the QD with respect to the patch, and the size of the patch. This protocol allows optimal control over all these parameters. In this study, we controlled all these parameters except QD orientation [31].

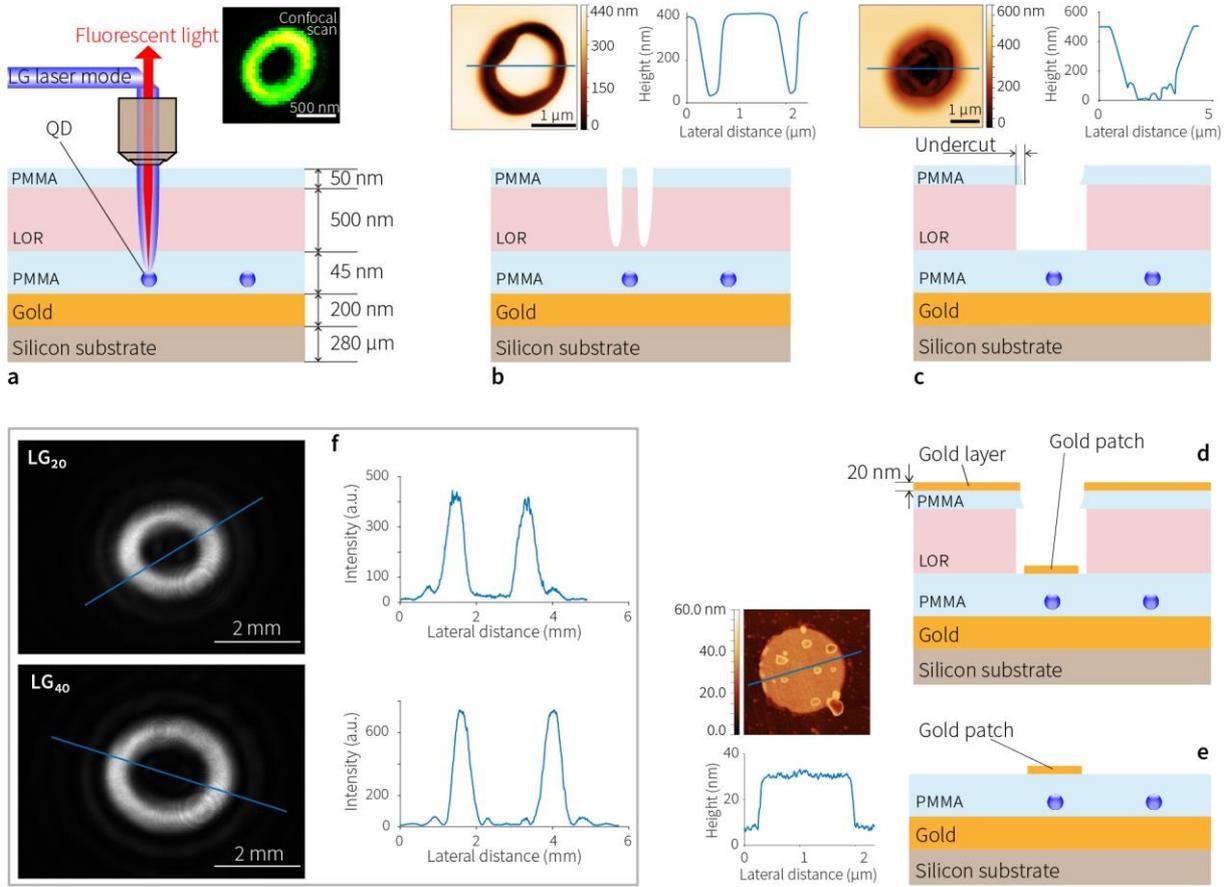

**Figure 2** *Deterministic in situ lithography on single-emitters using spatially modulated light*

Figures **a** to **e** illustrate the steps of the lithography protocol. The image above **a** is a confocal scan of a QD. The AFM image and the corresponding height profile above **b** depict a cylinder etched into the resist bi-layer, which is removed by chemical etching at step **c** as illustrated by the AFM image above it. The AFM image on the left of **e** shows a 20 nm high circular antenna patch of diameter 1.6 µm obtained after lift-off. **f** Cross-section intensity maps of LG$_{20}$ and LG$_{40}$ laser beams imaged by a camera, and the corresponding intensity profiles.

**Single photon emission from antennas under low excitation**

In homogeneous PMMA medium and under low intensity off-resonance excitation (405 nm laser pulsed at 2.5 MHz), a QD from the studied batch typically emitted single photons with a monoexponential decay of characteristic lifetime of about $\tau = 90$ ns (Figure 1c). Coupling a QD to the antenna changes its decay rate (Figure 3) and its single photon emission characteristics.

Most of the spontaneous decay response curves of these antennas can be fitted with a biexponential function with decay constants $\tau_{\text{fast}}$ (multiexciton) and $\tau_{\text{slow}}$ (exciton). Removing photon events corresponding to the fast decay through post-processing leaves mainly exciton recombination events, which are single photon emission events and therefore show lower $g^2(0)$. This is demonstrated in Figures 3d, e, where filtering out the fast component of the antenna decay (blue trace), leaves mostly exciton events (orange trace), which exhibits a negligible zero delay peak in the photon coincidence curve (Figure 3e, low $g^2(0)$ curve), thus validating the above hypothesis about the studied systems.

An excited QD decays in a cascade from the multiexciton states to the exciton state and finally to the ground state as depicted in Figure 3f. Auger processes, which are non-radiative and quench multiexciton emission, only influence multiexciton transitions such as biexciton to exciton, and are included in the multiexciton to exciton rate $\Gamma_{MX} = \Gamma_{MX}^{non-Auger} + \Gamma_{MX}^{Auger}$, where $\Gamma_{MX}^{Auger}$ and $\Gamma_{MX}^{non-Auger}$ denote the Auger and non-Auger multiexciton rates. Auger processes are inherently less efficient in these relatively large QDs due their size [32]. Purcell effect changes QD decay characteristics when it is coupled to the antenna. As the Purcell effect acts only on non-Auger electromagnetic transitions, Auger processes become even more ineffective in an antenna [25, 33]. This makes these antennas prone to multiexciton emission as demonstrated by the increased $g^2(0)$ (>0.25) in Figures 3b, c. Purcell factor $F_P$ can be calculated by comparing the QD exciton decay rate in a homogeneous medium $\Gamma_X$ to its exciton decay rate inside an antenna $\Gamma_X^{Antenna}$, where $\Gamma_X^{Antenna} = F_P \Gamma_X$. The values of $\Gamma_X$ and $\Gamma_X^{Antenna}$ are estimated respectively from $\tau_{slow}$ of the fitting curves. Using the typical emission lifetime of a QD in homogeneous medium, the $F_P$ of the three single photon source antennas of Figures 3a, b, c was noted as 38, 10, and 19, resp. Figure 3a is the emission response of the antenna of Figure 2e (AFM image), and Figure 4c shows its directive radiation pattern.

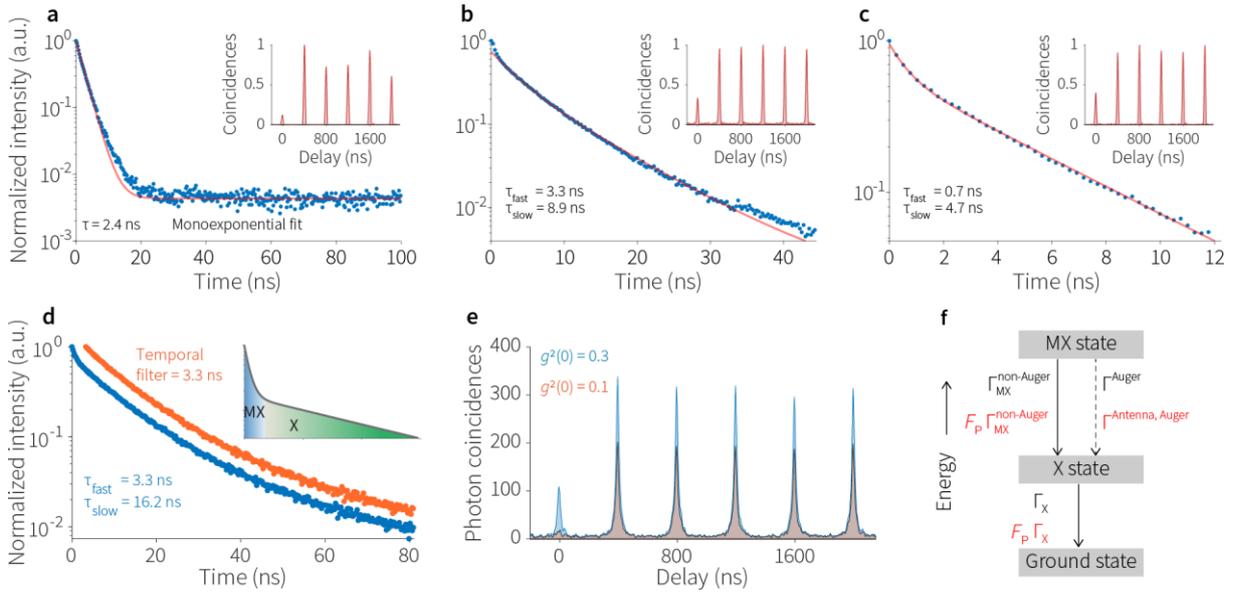

**Figure 3** *Antenna photoluminescence decay and photon correlation*

Fitted time-resolved photoluminescence responses and decay times of three antennas fabricated using in situ optical lithography with spatially modulated light **a–c**. The inset plots show the corresponding normalized coincidences, which signify $g^2(0)$. **d** shows the complete decay of an antenna (blue curve) and its post-processed decay after removing events < 3.3 ns (orange curve), and **e** shows the corresponding second order photon correlation curves—complete decay (blue) and filtered decay (orange). The inset in **d** is a simulated illustration of multiexciton (MX) and exciton (X) recombination in a biexponential decay. **f** Energy level diagram illustrating the transitions in QD decay—solid and dashed lines represent radiative and non-radiative transitions, resp.

The distinct rapidity of multiexciton emission allows for temporal filtering via post processing as shown in Figure 3d, e. The possibility of temporal filtering can be exploited to make such antennas better single photon sources [34]. As long as the exciton quantum yield of the QD is stable, the antenna emission can be switched very rapidly between single photon and multiphoton emission by varying the excitation intensity—this effect can find application in optoelectronic technologies [35].

**Emission enhancement under low excitation**

The widefield fluorescence images of Figure 4a demonstrate the enhancement in brightness due to a patch antenna. Both the images were acquired under similar conditions using continuous wave mercury lamp illumination at 438±12 nm. The top image of Figure 4a was collected after removing the resist bi-layer above the QD (at step c of Figure 2). We note that the QD fluorescence signal collected with this acquisition time is quite low. The luminescing edges are a result of laser and chemical etching. Positioning the plasmonic patch on this QD, and carrying out the required lift-off, made the fluorescence signal 6 times brighter (Figure 4a, bottom image). Under low intensity excitation, this bright antenna emitted at 640±33 nm (Figure 4b). The enhancement in fluorescence can be attributed to mainly two effects: 1) Plasmonic acceleration of recombination rates (generally, smaller patch antennas cause higher acceleration), and 2) increase in excitation and collection efficiency due to the antenna (generally, larger patch antennas exhibit more directional emission). It is noteworthy that under very low pulsed laser excitation intensity of only 15 mW/cm², we detected $1.7 \times 10^4$ photons/s with $g^2(0) = 0.2$. The remarkably high brightness of the antenna despite a moderate $F_P \approx 3$ can be attributed to the high collection of the photons emitted by it due to the directivity of the antenna and to the efficient excitation of the QD. This latter will be discussed in detail in the next section. The antenna patch directs emission very effectively—this is evident from highly directional radiation patterns of two antennas (Figures 4c, d) measured in the far field by Fourier plane imaging, which is in agreement with the simulation results for similar patch sizes [10]. The symmetry of the lobes in the polar plot depicts that the QD was positioned at the center with respect to the circular patch within 5 nm, which demonstrates the lithographic precision of this technique.

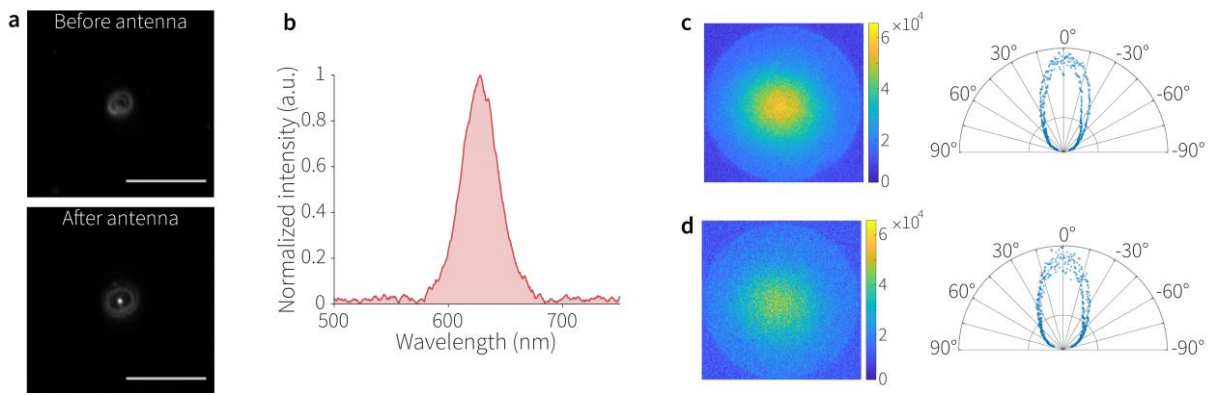

**Figure 4** *Fluorescence enhancement due to plasmonic nanoantennas*

**a** Widefield fluorescence images of the same QD before and after plasmonic patch deposition. Both images were acquired in grey scale by a charge coupled device camera in 200 ms and under similar excitation and collection conditions. The scale line represents 10 µm. **b** Emission spectrum of this antenna. Radiation patterns and the corresponding ($\vartheta$, $\varphi$) polar plots of two patch antennas showing their directional emission **c, d**.

**Antenna emission under increasing excitation**

Time-correlated single photon counting on a Hanbury Brown-Twiss optical measurement setup reveals that the emission characteristics of these antennas change considerably as the excitation intensity is varied. As these antennas involve plasmonically coupled large QDs, they are prone to multiexciton emission [25]. Exciting the bright antenna of Figure 4a more intensely resulted in a larger contribution of multiexcitons to its emission. The low average power of the pulsed laser (405 nm laser at 2.5 MHz in fundamental mode) required to excite this antenna demonstrate its efficiency (Figure 5a) and its sensitivity to nonlinear emission transition. At low power excitation (<0.5 nW), the antenna emission is mainly due to excitons and its $g^2(0)$ remains around 0.3 as shown by in the graph corresponding to P1 (Figure 5a, b), and curve follows a typical $1 - e^{-\beta P}$ trend, where $P$ is the average power of the laser and β is some coefficient (here β = 9.2 nW$^{-1}$). However, around point P2 and beyond, the emission becomes highly nonlinear and the trend was noted as $P^\alpha$, where $\alpha = 5.3$ in this case. Analyzing QD emission as a function of pump intensity, Zhang et al. have reported a very similar trend [36]; here we observe higher nonlinearity and a larger absorption cross-section due to the antenna. Plasmonic interactions provide efficient ways of trapping and propagating light [37, 38, 39], which increases the effective absorption cross-section of these antennas [40, 41]. Our measurements reveal a 1000-fold increase in the absorption cross-section due to the antenna.

We calculated the absorption cross-section of the antenna using the fitting parameter β as $\sigma_{\text{abs}} = A_{\text{spot}} \frac{hc}{\lambda} f_{\text{rep}} \beta$, where $A_{\text{spot}}$ is the area of the laser spot focused by the objective, $\frac{hc}{\lambda}$ is the excitation photon energy ($h$: Planck's constant, $c$: speed of light in vacuum, λ: laser wavelength), $f_{\text{rep}}$ is the repetition rate of the laser. Interestingly, the antenna above the QD increases its effective absorption cross-section to $4 \times 10^{-11}$ cm². When compared to $\sigma_{\text{abs}}$ of a QD (typically measured about $1 - 6 \times 10^{-14}$ cm², see Supplementary Information), this is a very significant increase, and it manifests in the very low power required to excite the antenna in the linear regime and to switch it to the nonlinear emission regime. This huge increase in the absorption cross-section, which leads to efficient excitation of the antenna, explains its high brightness (Figure 4a) despite its moderate $F_P \approx 3$ (Figure 5b).

From P2 to P4, $g^2(0)$ increases from 0.5 to 1 showing stronger multiexciton emission, and the lifetime reduces significantly because multiexciton recombination is considerably faster. This nonlinear emission suggests a superradiance effect [42, 36], where localized dipolar emitters emit collectively. Although the antenna contains a single QD, the multiple states created in the QD under higher excitation can be thought of as interacting dipolar emitters localized in it similar to coupled emitters in Dicke superradiance [43]. Another way to view this superradiance effect is as a plasmonic Dicke effect [44], where the cooperative emission from the antenna is due to the resonant energy transfer between the emitter and the plasmons rather than between multiple emitters. Such nonlinearity in emission introduced at a slightly higher power results in a substantial increase in brightness and recombination rate, thus creating an efficient and fast photon source.

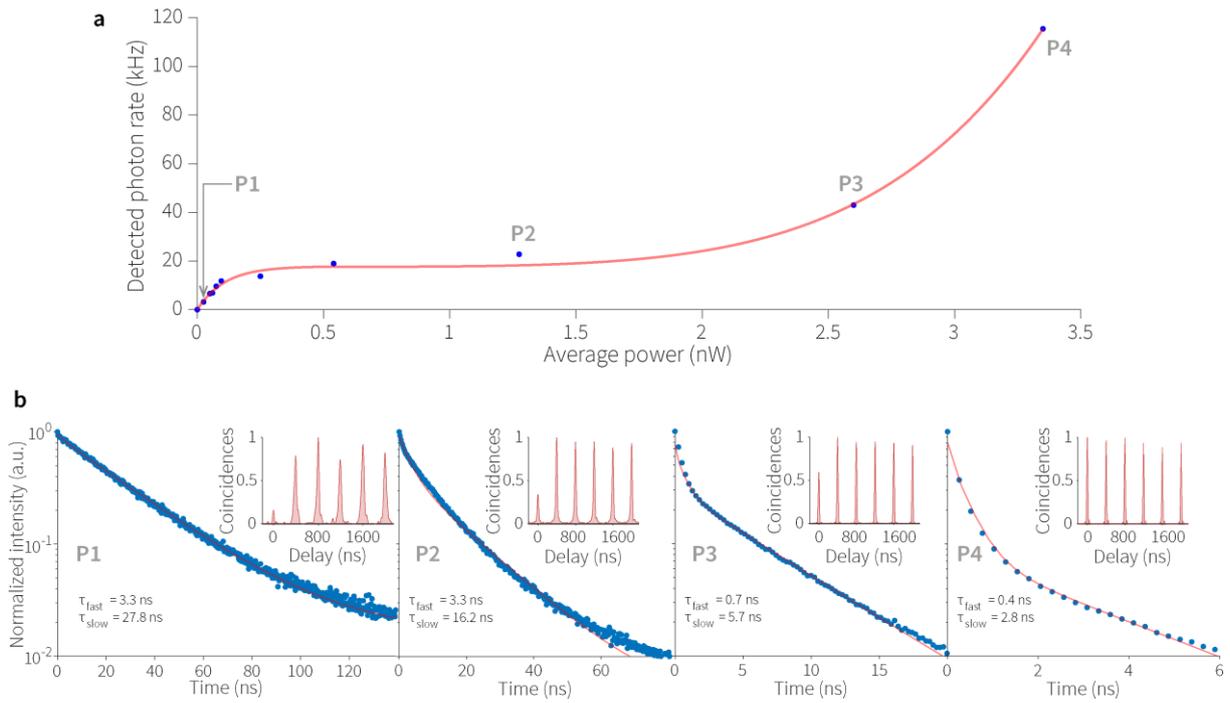

**Figure 5** *Exciton and multiexciton emission in plasmonic antenna*

**a** Detected photon intensity vs excitation power displays the transition from exciton emission to multiexciton emission (**b**), which is further highlighted by the emission decay and the normalized photon correlation curves for points P1–P4.

## Conclusion

The described deterministic nanolithography protocol helped us realize single-emitter plasmonic patch antennas with very high control over emitter positioning inside the antenna. The fabricated antennas showed directional and rapid single photon emission, and demonstrated nonlinear increase in brightness due to the high optical confinement inside the antenna and the high electron-hole pair confinement inside the nanoemitter. The dramatic increase in the absorption cross-section noted in plasmonic antennas opens opportunities for engineering these effects to create efficient light nanosources. Further investigation into the factors behind this effect, and controlling these factors by precise lithography will allow ways to overcome plasmonic losses, and create bright and fast emission sources. Moreover, a high absorption cross section is boon to applications that rely on light absorption such as solar energy conversion. In our future studies, we will investigate if the nonlinear emission of such patch antennas present opportunities that will benefit quantum technologies.

## Materials and Methods

**Sample preparation:** The CdSe/CdS core/shell QDs were synthesized chemically using the method detailed in the Supplementary Information. The sample substrate was a 0.28 mm thick polycrystalline Si wafer on which a 200 nm thick layer of Au (refractive index $n = 0.05 + i3.82$ at 630 nm after deposition measured by ellipsometry) was grown by vapor deposition using an adhesion layer of Cr (10 nm). A 10 nm thick film of PMMA was spin-coated on the Au film using a solution of 0.5% [m/m] PMMA (average molar mass = 101000) in toluene at 4000 rpm for 40 s. The sample was heated at 150°C for 2 minutes to evaporate the solvent from the polymer film. QDs dispersed in hexane were spin-coated on the PMMA film at 4000 rpm for 40 s to obtain well-distributed individual QDs. To sandwich the QDs in a dielectric medium, another layer of 1.5% [m/m] of PMMA solution in toluene was spin-coated on the QDs at 4000 rpm for 40 s. The sample was baked at 150°C for 2 minutes. The refractive index of the PMMA matrix film was measured by ellipsometry as $n = 1.50$ at 630 nm. For laser etching on the sample, a resist bi-layer was deposited. Initially, lift off resist LOR®5A was spin-coated at 7000 rpm for 40 s and the sample was baked at 150°C for 2 minutes to obtain a film of thickness 450 nm, and then a 10 nm thick layer of PMMA was spin-coated using a solution of 0.5% [m/m] PMMA in toluene. Finally, the sample was baked at 150°C for 2 minutes.

**Lithography**: The protocol is detailed in the paper.

**Sample characterization**: An *Olympus IX71* microscope equipped with a 0.8*NA* air objective (*Olympus LMPlanFL-100x*) was used to view to sample, which was mounted on a *PI P-713 XY Piezo Scanner* nanopositioning stage. Fluorescence widefield imaging arrangement used Hg lamp (*Olympus USH-1030L*) light filtered at 438±12 nm and the images were captured by a camera (*Photonic Science CCD*). Using confocal microscopy, the emitter was excited with a 405 nm pulsed laser (*PicoQuant LDH series* with 50 ps temporal width and operating at 2.5 MHz) to measure the photon statistics, and the lithography was performed by a spatially modified 473 nm CW diode laser (*LY473III-100*). Time-resolved photoluminescence and time-correlated single photon counting data was obtained after filtering the emission spectrally by a 630±46 nm bandpass filter (*Semrock 630/92 nm BrightLine®*) and spatially by a 150 μm pinhole, and using a *PicoHarp300* photon counting module and two *Micro Photon Devices PDM series* single photon avalanche photodiodes in a Hanbury-Brown and Twiss configuration. A 0.95*NA* air objective (*Olympus MPLAPON 100x*) was used to record the far-field emission pattern of the antennas by Fourier plane imaging with an EMCCD (electron multiplied charged coupled device) camera (*Andor iXon Ultra 897*). Tapping mode AFM measurements were performed using *Veeco Dimension 3100* AFM with *Bruker RTESPA* (Model: *MPP-11120-10*) tips. *Joel F2020* transmission electron microscope was used for imaging the QDs.

## Acknowledgements

The authors thank Jean-Paul Hugonin for insightful discussions, Willy Daney de Marcillac for spectrometric measurements, Loic Becerra and Stéphan Suffit for vapor deposition, and Bruno Gallas for ellipsometry experiments. This work was supported by DIM NanoK funding through the project PATCH and by ANR DELIGHT.

**Conflict of interests**

The authors declare no conflict of interests.

**Author contributions**

A.R.D. and A.M. conceived and developed the fabrication protocol, A.R.D. carried out the optical characterization, M.N. and B.D. synthesized the QDs, A.R.D., A.M., and Z.W. investigated the optical characterization data, and A.R.D and A.M. wrote the paper.

**Supplementary information accompanies the manuscript.**